\documentclass[apjl]{emulateapj}

\shorttitle{Morphologies of AGN hosts at z$\sim$2}
\shortauthors{Fan et al.}

\begin{document}

\newcommand{\kms}{\>{\rm km}\,{\rm s}^{-1}}
\newcommand{\reff}{r_{\rm{eff}}}
\newcommand{\msol}{M_{\odot}}
\newcommand{\gf}{{\tt GALFIT}~}
\newcommand{\mh}{H_{\rm{F160W}}}
\newcommand{\ser}{S\'ersic~}
\newcommand{\sext}{{\tt SExtractor}~}

\title{STRUCTURE AND MORPHOLOGY OF X-RAY SELECTED AGN HOSTS AT $1<Z<3$ IN CANDELS-COSMOS FIELD}

\author{Lulu Fan \altaffilmark{1,2,3}, Guanwen Fang\altaffilmark{4,5},Yang Chen\altaffilmark{6,1}, Jinrong Li\altaffilmark{1,2}, Xuanyi Lv\altaffilmark{1,2},Kirsten K. Knudsen\altaffilmark{3},Xu Kong\altaffilmark{1,2}}
\altaffiltext{1}{Center for Astrophysics, University of Science and Technology of China, 230026 Hefei, China; llfan@ustc.edu.cn}
\altaffiltext{2}{Key Laboratory for Research in Galaxies and Cosmology, USTC, CAS, 230026, Hefei, China}
\altaffiltext{3}{Department of Earth and Space Sciences, Chalmers University of Technology, Onsala Space Observatory, SE-439 92 Onsala, Sweden}
\altaffiltext{4}{Institute for Astronomy and History of Science and Technology, Dali University, Yunnan,671003, China}
\altaffiltext{5}{Key Laboratory of Modern Astronomy and Astrophysics, Nanjing University, Ministry of Education, Nanjing 210093, China}
\altaffiltext{6}{Astrophysics Sector, SISSA, Via Bonomea 265, 34136 Trieste, Italy}

\begin{abstract}
We analyze morphologies of the host galaxies of 35 X-ray selected active galactic nucleus (AGNs) at $z\sim2$ in the Cosmic
Evolution Survey (COSMOS) field using Hubble Space Telescope/WFC3 imaging taken from the Cosmic Assembly
Near-infrared Deep Extragalactic Legacy Survey (CANDELS). We build a control sample of 350 galaxies in total, by selecting ten non-active galaxies
drawn from the same field with the similar stellar mass and redshift for each AGN host.  By performing two dimensional fitting
with GALFIT on the surface brightness profile, we find that the distribution of \ser index (n) of AGN hosts does not show
a statistical difference from that of the control sample.  We measure the nonparametric morphological
parameters (the asymmetry index A, the Gini coefficient G, the concentration index C and the $M_{20}$ index) based on point source
subtracted images. All the distributions of these morphological parameters of AGN hosts are consistent with those of the control sample.
We finally investigate the fraction of distorted morphologies in both samples by visual classification. Only $\sim 15\%$
of the AGN hosts have highly distorted morphologies, possibly due to a major merger or interaction.
We find there is no significant difference in the distortion fractions between the AGN host sample and control sample.
We conclude that the morphologies of X-ray selected AGN hosts are similar to those of nonactive galaxies and most AGN activity is
not triggered by major merger.
\end{abstract}


\section{Introduction}

There is a long-standing question about what triggers AGN activity.
Even up till now, fifty years after quasars (QSOs, i.e. luminous AGNs) have been
discovered (Schmidt 1963), there is no convincing answer yet.
Major merger is a preferred candidate under the framework of hierarchical
structure formation (e.g. Hopkins et al. 2006). Other secular mechanisms, such as
minor merger, disk/bar instabilities, colliding clouds, supernova explosions,
have also been proposed to fuel black hole accretion and trigger
AGN activity (see Kormendy \& Kennicutt 2004; Martini 2004; Jogee 2006 for reviews).

One possible approach is to analyze morphologies of AGN host galaxies and
compare them with those of non-active galaxies. However, most morphological
analysis of AGN hosts can be seriously biased if AGNs contribute significantly
the total flux (e.g. Gabor et al. 2009; Pierce et al. 2010; B$\ddot{o}$hm et al. 2013).
A high-resolution image and a careful decomposition of a point-like source and AGN host will be necessary.
Until recently, with the advent of Hubble Space Telescope (HST), such morphological
analyses for large AGN samples at $z<1.3$ and $z\sim2$ in the rest-frame optical have been taken with
Advanced Camera for Surveys (ACS) and Wide Field Camera 3 (WFC3), respectively.

Previous studies have found different results. At low redshift ($z<1.3$),
some works found a significant fraction of AGNs have bulge-dominated morphologies from
a non-parametric classification (e.g. Grogin et al. 2005; Pierce et al. 2007), which
could be biased by the point-source component which is not removed in these studies.
While other works with a careful point-source removal found that AGN hosts show a wide range of
morphologies between bulge and disk dominated (Gabor et al. 2009) or over half of them have a
significant disk component (Cisternas et al. 2011).
In these studies, no enhancement of merger or interaction signatures have been found in
AGN hosts by comparing them with control samples.
Urrutia et al. (2008), however, found a high merger fraction (11 out of 13
in their luminous red QSOs sample). This may lead to an explanation that merger
fraction is dependent on AGN luminosity (Treister et al. 2012, but see also
Villforth et al. 2013).
At $z\sim2$, AGN hosts  seem to be disk-dominated with a similar merger fraction as those of nonactive galaxies
(Schawinski et al. 2011,2012; Simmons et al. 2012) or have a significant fraction
of disk galaxies with bulge component (Kocevski et al. 2012).

In this Letter we take the comprehensive structural and morphological analyses of X-ray selected
AGN host galaxies in the CANDELS-COSMOS field using HST/WFC3 H-band imaging  at
$1<z<3$. We try to find whether there is a special morphological type
which AGN hosts prefer to belong to. And we examine whether there
is a significant difference of 
morphological properties between AGN hosts and control samples.
In section 2, we describe our sample selections of X-ray selected AGN and control samples.
In section 3, we describe the methods to measure the morphological parameters. We at first perform
two dimensional light profile fitting with GALFIT using a \ser model representing
host galaxy contribution and a PSF model representing the point source contribution.
Then based on the point source subtracted images, we compute the nonparametric morphological
parameters. Finally we perform the visual classification based on the H-band image and
the residual image. In section 4, we give the main results of our morphological analyses.
In section 5, we summarize and discuss our results.  Throughout this letter, we assume
a concordance $\Lambda$CDM cosmology with $\Omega_{\rm m}=0.3$, $\Omega_{\rm \Lambda}=0.7$, $H_{\rm 0}=70$ $\kms$ Mpc$^{-1}$.

\section{Sample selection}

We use the X-ray selected AGN catalog presented in Bongiorno et
al. (2012). They selected AGNs from the XMM-COSMOS catalogue ( Cappelluti et
al. 2009)  with the optical identifications and multiwavelength
properties as discussed by Brusa et al. (2010) and new photometric
redshifts from Salvato et al. (2011). More than half of them
have available spectroscopic redshifts. Bongiorno et al. (2012) also
derived the host galaxy properties, such as colors and stellar masses,
based on a two-component model fit of AGN and host galaxy spectral
energy distributions (SEDs).

The central region of the COSMOS survey
(Scoville et al. 2007) has been imaged with HST/WFC3 as part of
CANDELS multi-cycle treasury programme (Grogin et al. 2011; Koekemoer
et al. 2011).  The HST WFC3/IR images have been prepared by drizzling
the individual exposures onto a grid with rescaled pixel sizes of 60
mas (Koekemoer et al. 2011).  We use the H-band (HST/WFC3 F160W filter) source catalogue presented in van der Wel et
al. (2012)  to match the X-ray selected AGN sample. With available
spectroscopic or photometric redshifts, we select 37 AGNs
with the H-band detections within the redshift range of $1<z<3$.
Among them, two sources lie on the edge of
CANDELS-COSMOS field and therefore have been discarded for further
morphological analysis. In total, we have 35 X-ray selected AGNs with
H-band images.
13 out of 35 sources have been classified as Type1 (unobscured) AGNs
and the rest 22 have been classified as Type 2 (obscured) according to
optical spectra, X-ray luminosities and multiwavelength SEDs (Salvato et al. 2011).
The high obscured fraction indicates that the light from the AGNs may be
not significant at the rest frame optical band where we do the morphological
analyses. 28 out of 35 sources have been detected in $2-10$keV band and the rest
has $0.5-2$keV band detection.
Their rest-frame $2-10$keV luminosities $L_{2-10keV}$ (without absorption correction) have
been derived using $L_{2-10keV}=4\pi D_L^2f_{2-10keV}(1+z)^{\Gamma-2}$,
where $f_{2-10keV}$ is the observed $2-10$keV flux, $D_L$ is
the luminosity distance and $\Gamma=1.8$ is the intrinsic AGN spectra index (e.g. Xue et al. 2011).
For seven objects with only $0.5-2$keV band detection, we derive their observed $2-10$keV flux
from $f_{0.5-2keV}$ by adopting an observed AGN spectra index $\Gamma_{obs}=1.4$ (e.g. Xue et al. 2011).
The $L_{2-10keV}$ of our sample have a range from $\sim3\times10^{43}$ erg s$^{-1}$
to $10^{45}$ erg s $^{-1}$ with a median value of $1.5\times 10^{44}$ erg s $^{-1}$,
about one order of magnitude higher than previous moderate-luminosity sample in
the similar redshift range (e.g. Simmons et al. 2012; Kocevski et al. 2012).
Stellar masses of host galaxies derived by SED fitting in
Bongiorno et al. (2012) show most of them are massive, with an average value
of $5\times10^{10}M_\odot$.

To construct the control sample, we randomly select ten non-active galaxies
from CANDELS-COSMOS H-band catalogue (van der Wel et al. 2012) for each AGN
host. Here we use the photometric redshifts of non-active galaxies derived by
Muzzin et al. (2013).We require that the selected non-active galaxies have similar stellar masses
and redshift as the matched AGN host. We select those non-active
galaxies with stellar masses within a factor of two and with redshift difference
less than 0.5 compared to AGN host (i.e. $|\Delta$ log $M_\star|\le 0.3$ and $|\Delta z| \le 0.5$).
Finally, 350 non-active galaxies in total meeting the criteria have been selected.

\begin{figure}
\epsscale{1.2}
\plotone{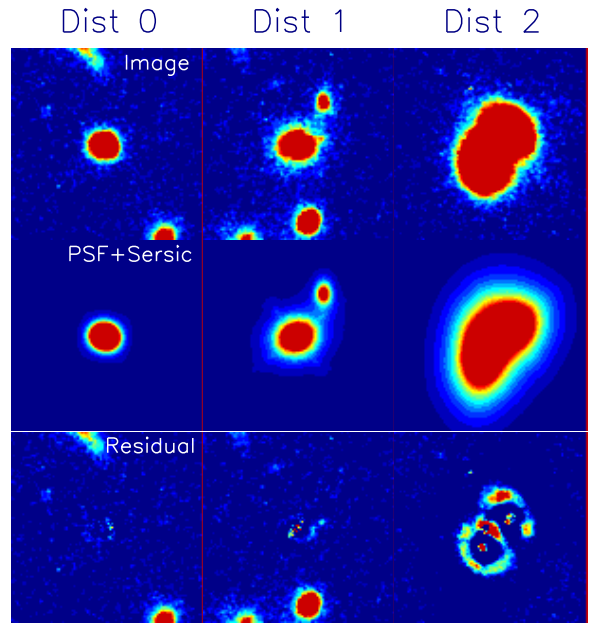}
\caption{Examples of our GALFIT analysis (see Section 3.1) and visual classification (see
Section 3.3). HST WFC3 F160W (H-band) images have been shown in the first row. And the corresponding
model (\ser + PSF) and residual images can be found in the second and third rows, respectively.
We also show examples of AGN host galaxy images arranged into three different distortion classes,
which are defined based on the H-band cutouts and residual images (see Section 3.3),
in three columns, respectively. }
\end{figure}

\section{Structural and Morphological parameters}

\subsection{GALFIT analysis}

We use the \emph{GALFIT} package  (Peng et al. 2002) to fit the surface brightness profiles
of our AGN host and control samples. The fits are performed in the
H-band cutouts images\footnote{http://www.mpia-hd.mpg.de/homes/vdwel/candels.html}.
As we did in our previous work (Fan et al. 2013a),
we use empirical PSF instead of model PSF. We extract
it from 43 stars with $S/N>50$ in the CANDELS-COSMOS field using PSFEx (Bertin 2011).

For the AGN host sample,
we use a PSF to model the nuclear point source, plus a \ser function to model
the host galaxy (See Figure 1). We constrain \ser index within a proper range (i.e. $0.1\le n \le 10$).
For those objects with nuclear point source dominant (6/35, defined by mag$_{host}>$ mag$_{(point\ source)}$), the two-dimensional fitting
with a \ser+PSF model will become very unreliable. We use a single PSF model instead for
these objects. For the control sample, we use the structural parameters from van der Wel et al. (2012)
which have been measured with a single \ser function.

\begin{figure*}
\begin{center}
\includegraphics[ width=0.49\textwidth]{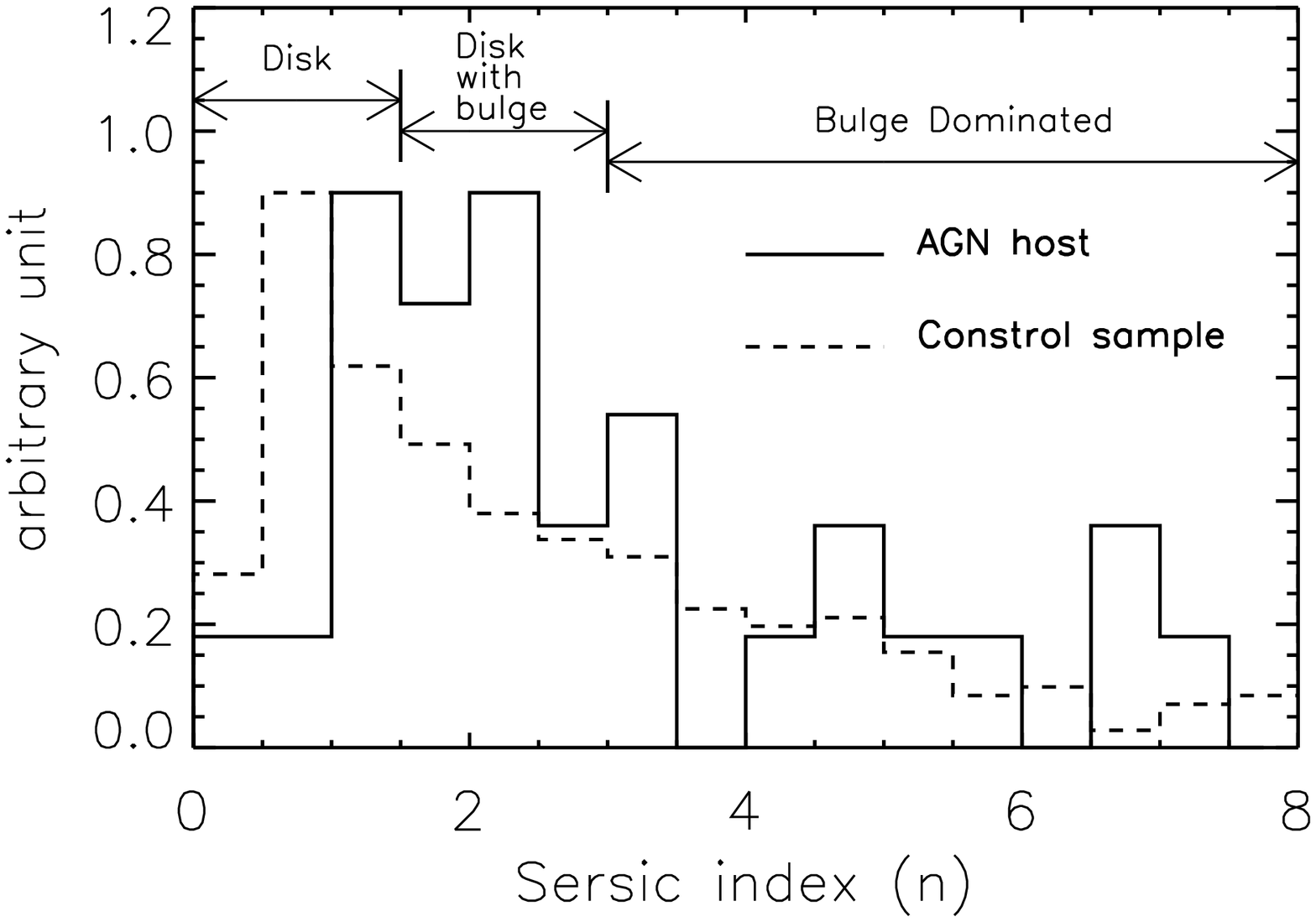}
\includegraphics[ width=0.49\textwidth]{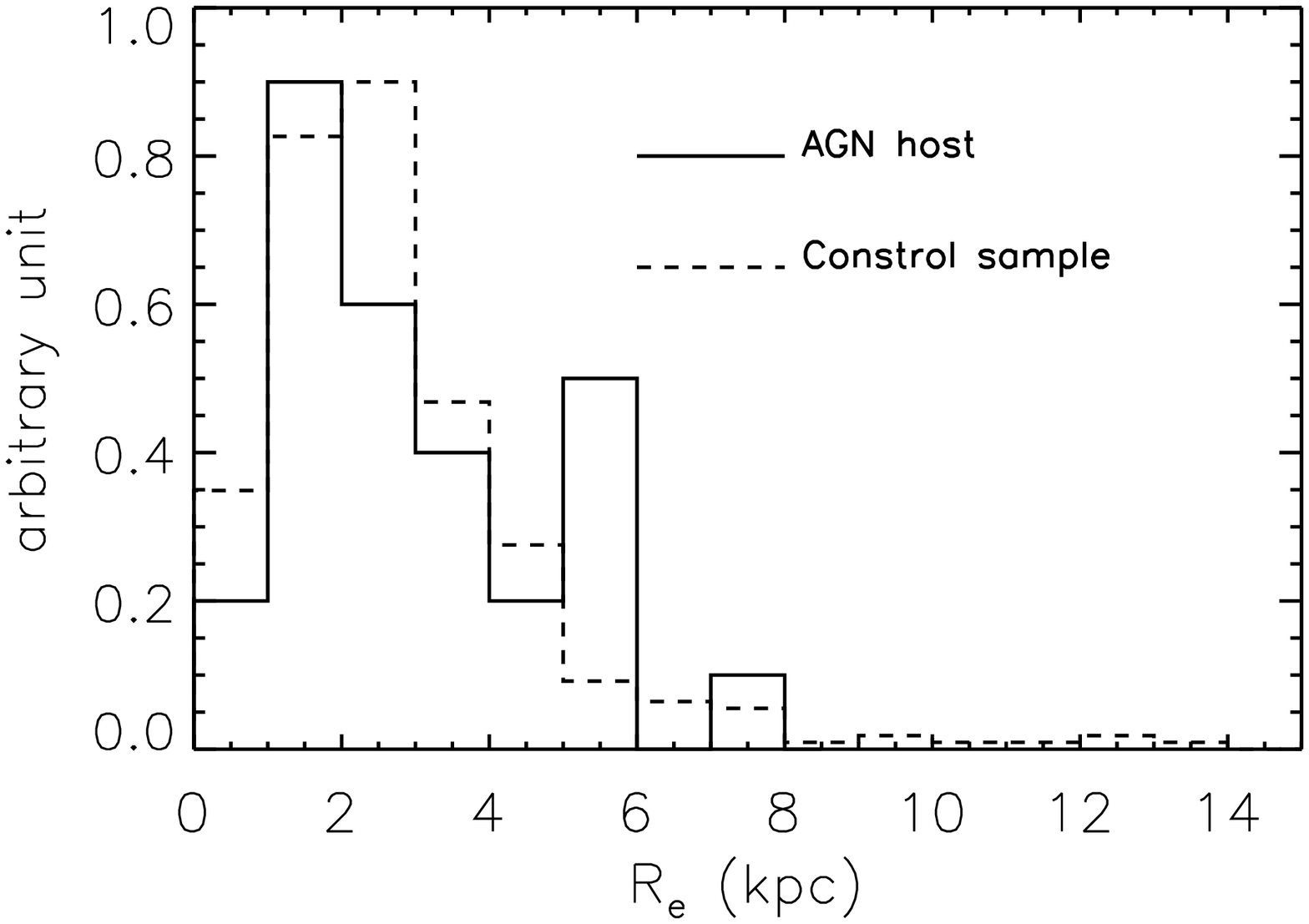}
\caption{Structural parameters of AGN host and non-active galaxies. \textit{Left:} the distribution of  \ser\ indices $n$. \textit{Right:} the physical effective radii $R_{\rm e}$. The solid line represents the X-ray selected AGN host galaxies at $z\sim2$, while the dashed line represents the control sample. The Kolmogorov-Smirnov (KS) test probabilities ($P_{KS}$) are 0.47 and 0.56 for the distributions of \ser\ indices and effective radii of the AGN host and control samples, respectively.  }
\end{center}
\end{figure*}

\subsection{Nonparametric Morphological Parameters}

For AGN host and control samples, we also measure nonparametric morphological
parameters, such as Gini coefficient (the relative distribution of the galaxy pixel flux values),
$M_{20}$ (the second-order moment of the brightest 20\% of the galaxy's flux),
concentration index (C) and rotational asymmetry index (A). Compared to \ser index, these parameters are
model-independent and therefore can be applied to irregulars, as well as standard Hubble-type galaxies
( e.g., Abraham et al. 1996; Conselice 2003; Lotz et al. 2004; Kong et al. 2009, Fang et al. 2009, 2012; Wang et al. 2012).

For non-active galaxies in control sample, we measure these parameters using the original
H-band images. While for AGN hosts, the case becomes a bit complicated.
The presence of the highly symmetric nuclear point source will strongly
bias all the measurements of host morphological parameters (Pierce et al. 2010; B$\ddot{o}$hm et al. 2013).
Therefore we measure these parameters of AGN host galaxies
with the same way as Gabor et al.(2009) did.
We subtract the best-fit model nuclear point source derived in Section 3.1 from each AGN H-band image. For objects
with nuclear point source dominant, we use residual images from our PSF-only-fit subtraction.

\subsection{Visual Classification}

Besides the structural and morphological parameters measurements of
AGN host and control sample, we also investigate the merger or interaction
fraction  of AGN host and control samples via visual classification.
Using the same method presented in Cisternas et al. (2011), all AGN host and non-active
galaxies have been visually classified by five human classifiers independently,
based on the H-band cutouts and residual images by subtracting the best-fit \ser+PSF model
presented in Section 3.1.  We use the same definition of ``distortion class" in Cisternas et al. (2011)
to describe the degree of distortion of the galaxy.
There are three classes used in total: Dist 0, Dist 1 and Dist 2.
Class ``Dist 0" represents  undisturbed and smooth galaxies, showing no interaction signatures.
Class ``Dist 1" represents  galaxies with mild distortions, possibly due to minor merger or accretion.
Class ``Dist 2" represents  galaxies with strong distortions, potential signs for ongoing
or recent mergers. Illustrative examples of distortion classes can be
found in Figure 1.

\section{Results}

In Figure 2, we plot the distributions of  \ser\ index $n$ (left panel) and the physical effective radii $R_{\rm e}$ (right panel)
of AGN host and non-active galaxies. 
The objects with a dominant nuclear point source are not included in this plot.
We compute the Kolmogorov-Smirnov (KS) test probabilities ($P_{KS}$) which
are 0.47 and 0.56 for the distributions of \ser\ indices and effective radii of the AGN host and control samples, respectively.
The results are consistent with  both \ser indices and effective radii of AGN hosts being drawn
from the same distributions as the control sample.

The distribution of \ser indices of X-ray selected AGN host galaxies indicates a broad range
of morphologies, from disk-dominated ($n<1.5$), disk with a prominent bulge component (1.5$\le$n$\le$3.0) to bulge-dominated
($n>3.0$). The fractions of disk-dominated, intermediate and bulge-dominated morphologies
are 27.6\%, 34.5\% and 37.9\%, respectively, in our AGN host galaxy sample.
Disks (with and without a prominent bulge component) are the most common morphology in
X-ray selected AGN host galaxies (close to two thirds of the entire sample),
while there are also a significant fraction (more than one thirds of the entire sample) dominated by bulge.
These fractions are quantitatively consistent with those of X-ray selected AGN host galaxies at redshifts
$0.3<z<1.0$ (Gabor et al. 2009).
However, some previous findings (e.g., Schawinski et al. 2011, Simmons et al. 2012) showed a higher
fraction of X-ray selected AGN host galaxies in disk galaxies at similar redshift $z\sim2$.
They found that  $\sim80-90\%$ of X-ray selected AGN host galaxies with luminosities of
$10^{42}$ erg s$^{-1}$ $<L_X<10^{44}$ erg s$^{-1}$ had low \ser indices ($n<3$) indicative of
disk dominated light profile.
Compared to their results, AGN hosts in our sample have a lower fraction of
disks and a higher fraction of bulge-dominated morphology.
One possible explanation of the difference between our results is that AGN host morphology
could vary with X-ray luminosity.  The X-ray luminosities of our sample
have an average value of $\sim10^{44}$ erg s$^{-1}$, about one order of magnitude higher than
those in their previous works. Host galaxies of AGNs with higher X-ray luminosities may have a higher fraction
of bulge-dominated morphology. This possibility has been tested and confirmed by an independent work of
Kocevski et al. (2012) via visual inspection of a moderate-luminosity AGN sample
at $z\sim2$.  They observed a dramatic rising in the bulge-dominated fraction:
from 18.4\% in the subsample with $L_X<10^{43}$ erg s$^{-1}$ to
40.6\% in the subsample with $L_X>10^{43}$ erg s$^{-1}$.
We notice that the bulge-dominated fraction ($\sim40\%$) in their subsample with $L_X>10^{43}$ erg s$^{-1}$
agrees well with that value in our result.

\begin{figure}
\epsscale{1.2}
\plotone{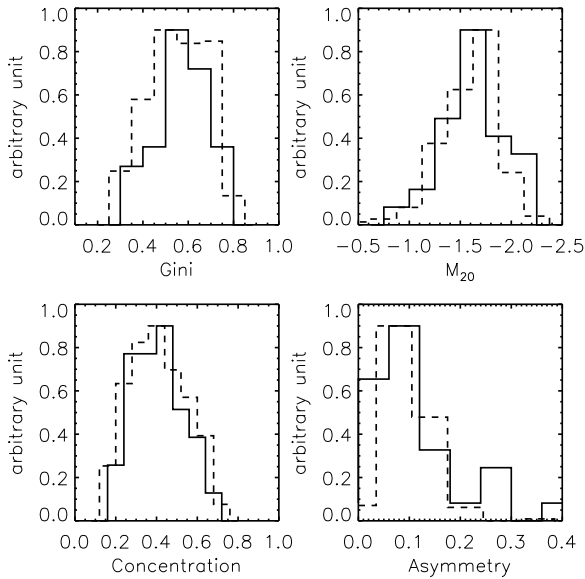}
\caption{The distributions of nonparametric morphological parameters (\textit{top-left:} Gini;
\textit{top-right:} $M_{20}$; \textit{bottom-left:} concentration; \textit{bottom-right:} asymmetry ) of AGN host and non-active galaxies.
 The solid line represents the X-ray selected AGN host galaxies at $z\sim2$, while the dashed line represents the control sample.
 KS test probabilities that nonparametric morphological parameters of AGN host galaxies are drawn from the same distributions
as their control sample are 0.91, 0.85, 0.94 and 0.38 for Gini, $M_{20}$, concentration and asymmetry, respectively. }
\end{figure}

In Figure 3, we plot the distributions of four nonparametric morphological parameters (Gini,
$M_{20}$, concentration and asymmetry ) of AGN host and non-active galaxies.
We perform a KS test to determine whether
the AGN host and control sample populations are consistent with being drawn from the
same underlying distribution. We find that all distributions of these parameters
for AGN hosts have no difference to those of non-active galaxies, with KS test
probabilities of 0.91,0.85,0.94 and 0.38 for Gini, $M_{20}$, concentration and asymmetry, respectively. We notice that the nonparametric morphological parameters measurements of six objects with nuclear point source dominant are very uncertain. However, we find that the inclusion of these six objects or not will not change our main conclusion.

\begin{deluxetable}{lcc}
\tabletypesize{\scriptsize}
\tablecaption{Mean distortion class classification of AGN host and control samples}
\tablewidth{0.4\textwidth}
\tablehead{
\colhead{Distortion Class} & \colhead{$\mu_{AGN}^a$} & \colhead{$\mu_{CS}^a$} }
\startdata
Dist-0  &  54.3\% $\pm$ 10.1\% & 57.1\% $\pm$ 6.9\%   \\
Dist-1  &  30.9\% $\pm$ 8.7\%  & 29.0\% $\pm$ 5.3\%   \\
Dist-2  &  14.8\% $\pm$ 2.4\%  & 13.9\% $\pm$ 4.3\%
\enddata
\tablenotetext{a}{Mean of the 5 classifications}
\end{deluxetable}

In Table 1, we summarize the mean fractions of three distortion classes in AGN host and control samples via
visual classification. Several main results can be addressed:
\begin{enumerate}
  \item Over 50\% of AGN host galaxies have undisturbed and smooth light profile,
showing no evidence for ongoing merger.
  \item Visual morphologies of near 50\% of AGN host galaxies are distorted to different degrees, showing
possible signatures for ongoing minor/major mergers. However, among these distorted AGN host galaxies,
most of them have mild morphological distortion. Only $<15\%$ of the entire AGN host galaxies show strong distortions,
which are potential signs for ongoing or recent major mergers.
  \item The fractions of distortion classes of AGN host galaxies are consistent with  those of the control galaxies.
We find no evidence that AGNs host a higher fraction of distorted morphologies (Dist 1 and Dist 2 classes) than non-active
galaxies.
\end{enumerate}
These results are generally consistent with the previous findings
of Cisternas et al. (2011) at lower redshift ($0.3<z<1.0$) and of Kocevski et al. (2012)
at similar redshift ($z\sim2$).


\section{Summary and Discussion}

In this letter, we analyze the structure and morphology of X-ray selected
AGN host galaxies in the CANDELS-COSMOS field using HST/WFC3 imaging in H band at
$z\sim2$ and compare them with those of a mass-matched non-active galaxy sample.
Our primary findings are as follows: 1) Near two thirds of X-ray selected AGN host galaxies in our
sample at $z\sim2$ have disk-like morphologies (including disk-dominated and disk with a bulge component
morphologies), while a significant fraction (over one thirds) of them are bulge dominated;
2) All structural (\ser index n and effective radius $R_e$) and morphological
(Gini, $M_{20}$,concentration and asymmetry) parameters of AGN host galaxies
in our sample have the similar distributions as those of non-active galaxies.
We conclude that, from the point of view of structure and morphology,
AGN host galaxies and non-active galaxies are indistinguishable.
From other point of view of such as color, color gradients and stellar population
properties, Rosario et al (2013) also found that X-ray selected AGN host galaxies
and non-active galaxies are indistinguishable. 3) Only a small fraction ($\sim 15\%$) of X-ray selected AGN host galaxies have
major merger signatures according to our visual classification. Compared to
the mass-matched non-active galaxy sample, AGN hosts do not show a significant excess
of distorted morphologies.

As shown by previous host galaxy simulations (e.g., Simmons \& Urry 2008; Gabor et al. 2009; Pierce et al. 2010),
the result that AGN host galaxies are disk-dominated characterized by a low fitted \ser index n is reliable.
While disk contribution cannot be ruled out in bulge-dominated galaxy with a high fitted \ser index n.
Simmons \& Urry (2008) found that bulge-dominated AGN host galaxies  with $n>4$
may have a significant disk contribution (up to 45\% of their total host galaxy light).
The disk fraction ($\sim63\%$) of AGN host galaxies in our sample should be taken as a lower limit.
This value is a bit lower than the disk fraction $75\%$ in Simmons et al. (2012),
and is much lower than the disk fraction $\sim90\%$ in Schawinski et al. (2011).
The relatively lower disk fraction in our sample could be the result of different sample selections.
The X-ray luminosities of AGNs in our sample are on average one order of magnitude higher
than those in Schawinski et al. (2011).

Complemented with several previous findings at $0<z<3$
(e.g., Gabor et al. 2009, Cisternas et al. 2011, Schawinski et al. 2011, Simmons et al. 2012, Kocevski et al. 2012),
some consistent results can be summarized for X-ray selected AGNs with $10^{42}$ erg s$^{-1}$ $<L_X<10^{45}$ erg s$^{-1}$  up to $z\sim3$:
disk morphologies ; no enhanced distorted morphologies and similar structure and morphology as non-active galaxies.
These indicate that major mergers are not necessary for triggering AGN activities in the galaxies of the X-ray selected sample.
Secular internal processes, such as gravitational instabilities  and dynamical friction,
should play a crucial role in triggering X-ray selected AGN activities and black hole-host galaxy co-evolution.
However, our findings do not conflict with supermassive black hole (SMBHs)-host galaxy co-evolution  scenarios in which
major mergers are responsible for triggering both star formation and luminous QSO activities (e.g., Hopkins et al. 2006).  Unlike X-ray selected AGNs, luminous QSOs commonly have very high fraction (up to $\sim 100\%$) in major merger and are specifically in dust-shrouded, merger-induced starbursts (e.g., Urrutia et al. 2008; Treister et al. 2012). Thus two different modes of BH triggering and growth have been indicated: the BH growth by secular evolution of disk galaxies is important for low-luminosity AGNs, while major mergers may feed luminous quasars (e.g., Kormendy \& Ho 2013).



\acknowledgments

We thank the referee for the careful reading and the valuable comments that helped improving our paper.
We specially thank the CANDELS and COSMOS teams for making their excellent data products publicly available.
This work was also supported by the National Natural Science Foundation of China
(NSFC, Nos. 11203023, 11225315, 11320101002 and 11303002), the Open Research
Program of Key Laboratory for the Structure and Evolution of Celestial
Objects, CAS, the Specialized Research Fund for the Doctoral Program of
Higher Education (SRFDP, No. 20123402120015, 20123402110037) and Chinese Universities
Scientific Fund (CUSF). LF and KK acknowledge the Knut and Alice Wallenberg Foundation for support.

\textit{Facilities}: HST (WFC3); XMM-Newton


\begin{thebibliography}{99}

\bibitem[Abraham et al.(1996)]{1996MNRAS.279L..47A} Abraham, R.~G., Tanvir, N.~R., Santiago, B.~X., et al.\ 1996, \mnras, 279, L47
\bibitem[Bertin(2011)]{Bertin2011} Bertin, E.\ 2011,Astronomical Data Analysis Software and Systems XX, 442, 435
\bibitem[B{\"o}hm et al.(2013)]{2013A&A...549A..46B} B{\"o}hm, A., Wisotzki, L., Bell, E.~F., et al.\ 2013, \aap, 549, A46
\bibitem[Bongiorno et al.(2012)]{2012MNRAS.427.3103B} Bongiorno, A., Merloni, A., Brusa, M., et al.\ 2012, \mnras, 427, 3103
\bibitem[Brusa et al.(2010)]{2010ApJ...716..348B} Brusa, M., Civano, F., Comastri, A., et al.\ 2010, \apj, 716, 348
\bibitem[Cappelluti et al.(2009)]{2009A&A...497..635C} Cappelluti, N., Brusa, M., Hasinger, G., et al.\ 2009, \aap, 497, 635
\bibitem[Cisternas et al.(2011)]{2011ApJ...726...57C} Cisternas, M., Jahnke, K., Inskip, K.~J., et al.\ 2011, \apj, 726, 57
\bibitem[Conselice(2003)]{2003ApJS..147....1C} Conselice, C.~J.\ 2003, \apjs, 147, 1
\bibitem[Fan et al.(2013a)]{2013MNRAS.431L..15F} Fan, L., Chen, Y., Er, X., et al.\ 2013, \mnras, 431, L15
\bibitem[Fang et al.(2009)]{2009RAA.....9...59F} Fang, G.-W., Kong, X., \& Wang, M.\ 2009, Research in Astronomy and Astrophysics, 9, 59
\bibitem[Fang et al.(2012)]{2012ApJ...751..109F} Fang, G., Kong, X., Chen, Y., \& Lin, X.\ 2012, \apj, 751, 109
\bibitem[Gabor et al.(2009)]{2009ApJ...691..705G} Gabor, J.~M., Impey, C.~D., Jahnke, K., et al.\ 2009, \apj, 691, 705
\bibitem[Grogin et al.(2005)]{2005ApJ...627L..97G} Grogin, N.~A., Conselice, C.~J., Chatzichristou, E., et al.\ 2005, \apjl, 627, L97
\bibitem[Grogin et al.(2011)]{2011ApJS..197...35G} Grogin, N.~A., Kocevski, D.~D., Faber, S.~M., et al.\ 2011, \apjs, 197, 35
\bibitem[Hopkins et al.(2006)]{2006ApJS..163....1H} Hopkins, P.~F., Hernquist, L., Cox, T.~J., et al.\ 2006, \apjs, 163, 1
\bibitem[Jogee(2006)]{2006LNP...693..143J} Jogee, S.\ 2006, Physics of Active Galactic Nuclei at all Scales, 693, 143
\bibitem[Kocevski et al.(2012)]{2012ApJ...744..148K} Kocevski, D.~D., Faber, S.~M., Mozena, M., et al.\ 2012, \apj, 744, 148
\bibitem[Kong et al.(2009)]{2009ApJ...702.1458K} Kong, X., Fang, G., Arimoto, N., \& Wang, M.\ 2009, \apj, 702, 1458
\bibitem[Kormendy \& Ho(2013)]{2013ARA&A..51..511K} Kormendy, J., \& Ho, L.~C.\ 2013, \araa, 51, 511
\bibitem[Kormendy \& Kennicutt(2004)]{2004ARA&A..42..603K} Kormendy, J., \& Kennicutt, R.~C., Jr.\ 2004, \araa, 42, 603
\bibitem[Koekemoer et al. (2011)]{2011ApJS..197...36K} Koekemoer, A., Faber, S., Ferguson, H., et al.\ 2011, \apjs, 197, 36
\bibitem[Lotz et al.(2004)]{2004AJ....128..163L} Lotz, J.~M., Primack, J., \& Madau, P.\ 2004, \aj, 128, 163
\bibitem[Martini(2004)]{2004cbhg.symp..169M} Martini, P.\ 2004, Coevolution of Black Holes and Galaxies, 169
\bibitem[Muzzin et al.(2013)]{2013ApJS..206....8M} Muzzin, A., Marchesini, D., Stefanon, M., et al.\ 2013, \apjs, 206, 8
\bibitem[]{Peng02}{Peng, C., Ho, L., Impey, C., et al.} 2002, AJ, 124, 266
\bibitem[Pierce et al.(2007)]{2007ApJ...660L..19P} Pierce, C.~M., Lotz, J.~M., Laird, E.~S., et al.\ 2007, \apjl, 660, L19
\bibitem[Pierce et al.(2010)]{2010MNRAS.408..139P} Pierce, C.~M., Lotz, J.~M., Salim, S., et al.\ 2010, \mnras, 408, 139
\bibitem[Rosario et al. (2013)]{rosaro13} Rosario, D., Mozena, M., Wuyts, S., et al. \ 2013, \apj, 763, 59
\bibitem[Salvato et al.(2011)]{2011ApJ...742...61S} Salvato, M., Ilbert, O., Hasinger, G., et al.\ 2011, \apj, 742, 61
\bibitem[Schawinski et al.(2011)]{2011ApJ...727L..31S} Schawinski, K., Treister, E., Urry, C.~M., et al.\ 2011, \apjl, 727, L31
\bibitem[Schawinski et al.(2012)]{2012MNRAS.425L..61S} Schawinski, K., Simmons, B.~D., Urry, C.~M., et al.\ 2012, \mnras, 425, L61
\bibitem[Schmidt(1963)]{1963Natur.197.1040S} Schmidt, M.\ 1963, \nat, 197, 1040
\bibitem[Scoville et al.(2007)]{2007ApJS..172....1S} Scoville, N., Aussel, H., Brusa, M., et al.\ 2007, \apjs, 172, 1
\bibitem[Simmons \& Urry(2008)]{2008ApJ...683..644S} Simmons, B.~D., \& Urry, C.~M.\ 2008, \apj, 683, 644
\bibitem[Simmons et al.(2012)]{2012ApJ...761...75S} Simmons, B.~D., Urry, C.~M., Schawinski, K.,et al.\ 2012, \apj, 761, 75
\bibitem[Treister et al.(2012)]{2012ApJ...758L..39T} Treister, E., Schawinski, K., Urry, C.~M., et al.\ 2012, \apjl, 758, L39
\bibitem[Urrutia et al.(2008)]{2008ApJ...674...80U} Urrutia, T., Lacy, M., \& Becker, R.~H.\ 2008, \apj, 674, 80
\bibitem[van der Wel et al.(2012)]{2012ApJS..203...24V} van der Wel, A., Bell, E.~F., H{\"a}ussler, B., et al.\ 2012, \apjs, 203, 24
\bibitem[Villforth et al.(2013)]{2013arXiv1303.1874V} Villforth, C., Hamann, F., Koekemoer, A., et al.\ 2013, arXiv:1303.1874
\bibitem[Wang et al.(2012)]{2012ApJ...752..134W} Wang, T., Huang, J.-S., Faber, S.~M., et al.\ 2012, \apj, 752, 134
\bibitem[Xue et al.(2011)]{2011ApJS..195...10X} Xue, Y.~Q., Luo, B., Brandt, W.~N., et al.\ 2011, \apjs, 195, 10

\end{thebibliography}
\end{document}